\documentclass[preprint,12pt]{aastex}

\newcommand{\degree}{\mbox{$^{\circ}$}}

\newcommand{\massloss }{\mbox{$\dot M$}}

\newcommand{\lsun}{\mbox{L$_\odot$}}
\newcommand{\msun}{\mbox{M$_\odot$}}

\input{epsf}

\def\plotfiddle#1#2#3#4#5#6#7{\centering \leavevmode
\vbox to#2{\rule{0pt}{#2}}
\includegraphics{#1}}

\begin{document}

\title {\bf Multi-spectral Observations of Lunar Occultations: I. Resolving The Dust Shell Around AFGL 5440}
\author{Paul M. Harvey\altaffilmark{1} and Andrew Oldag\altaffilmark{1}
}

\altaffiltext{1}{Astronomy Department, University of Texas at Austin, 1 University Station C1400, Austin, TX 78712-0259;  pmh@astro.as.utexas.edu, feardrew@astro.as.utexas.edu}

\begin{abstract}

We present observations and modeling of a lunar occultation of the dust-enshrouded carbon star
AFGL 5440.  The observations were made over a continuous range of wavelengths from 1 -- 4\micron\
with a high-speed spectrophotometer designed expressly for this purpose.   We find
that the occultation fringes cannot be fit by any single-size model.  We use the DUSTY radiative
transfer code to model a circumstellar shell and fit both the observed occultation light curves
and the spectral energy distribution described in the literature.   We find a strong constraint
on the inner radius of the dust shell, T$_{max}$ = 950 K $\pm$ 50K, and optical depth at 5\micron\
of 0.5 $\pm$ 0.1.  The observations are best fit by models with a density gradient of $\rho \propto r^{-2}$
or the gradient derived by Ivezi\'c \& Elitzur for a radiatively driven hydrodynamic outflow.  Our models cannot fit
the observed IRAS 60\micron\ flux without assuming a substantial abundance of graphite or by assuming
a substantially higher mass-loss rate in the past.

\end{abstract}



\section{Introduction}

Mass loss from post-main-sequence stars provides a large fraction of the heavy element abundance and solid
particle content of the interstellar medium (e.g. Wallerstein \& Knapp 1998; Ferrarotti \& Gail 2006).   The mechanism(s) of the mass loss during the AGB phase
of stellar evolution are thought to involve both radiation pressure and stellar pulsation \citep{suh97,walknapp98,schroder98}, but 
most details of this process are not well understood.  Many uncertainties about these processes can
be clarified by studies of the spatial structure of circumstellar mass-loss shells.  For example,
a number of molecular line studies of the extended envelopes around AGB stars have found strong evidence for
periodic variations in mass-loss rates leading to the appearance of ``rings'' in the radial distribution
of molecular emission lines, (e.g. Fong, Meixner \& Shah 2003, Olofsson et al. 1996).  Very deep, sensitive imaging studies have also found similar phenomena in the dust around
the most nearby extreme example of these objects IRC+10216 \citep{mauron00}.  In order to study the inner regions of
these circumstellar shells, however, angular resolutions well under 1 arcsec are required.  For example,
at a distance of 1 kpc, the dust evaporation radius around a 10$^4$ \lsun\ star corresponds to an angular
radius of 20 milli-arcsec (mas).  Speckle interferometry and more recently adaptive optics observations have
enabled resolutions of order 0.1 arcsec \citep{hoff01,biller05}, while lunar occultation observations and
multi-aperture interferometry have pushed angular resolutions to the milli-arcsecond level, e.g. reviews by
\citet{quir04} and \citet{mon03}.

Until recently most interferometric observations have been made in
typically one or two relatively broad bands.  We present here observations of a lunar occultation
of the star AFGL 5440 (aka OH 06.86-1.5, IRAS 18036-2344),
made with a high-speed infrared spectrophotometer, pMIRAS \citep{harv03}, developed as a prototype for a more
ambitious instrument now nearing completion.  This star has been classified as a carbon-rich AGB star on the
basis of its IRAS LRS spectrum \citep{zuck86,volk89,kwok97}.  \citet{groen02} have estimated
the distance to be 2.25 kpc.  Near-infrared through far-infrared photometry of the source has been summarized
by \citet{gug93} and more recently by \citet{guand06}, including a combination of ground based photometry, the IRAS values, and more
recent MSX results.  The reported distance and photometry imply a luminosity of 1.4$\times$10$^4$ \lsun.

In addition to the observations of AFGL 5440, we
also discuss observations that we have made 
of two ``calibration'' stars 
in order to
understand the limitations of our observation/analysis process.  These objects are cool stars with no
detectable circumstellar dust shell based on their near-infrared and IRAS colors, IRC+00233 (M7) and HD 155292 (K2).  

Our observations cover the entire 1 - 4\micron\ spectral region
with a resolving power 
that varies from $\sim$ 20 at the shortest wavelengths to $\sim$ 100 at the long end.  Our time resolution of 8 msec permits an effective angular resolution of a
couple milli-arcsec, with the exact resolution being a strong function of the signal-to-noise ratio as discussed
later.
The broad wavelength coverage allows us to observe simultaneously the Fresnel fringe pattern
of the obscured central star at the shorter wavelengths together with the circumstellar dust emission from the
warmest part of the dust distribution.  In \S \ref{obs} we describe the details of the observations and instrumental
parameters and the basic data reduction process.  Then in \S \ref{model} we describe
various ways we have modelled the star$+$shell in order to determine the limits on the circumstellar shell
structure placed by our observations.  Finally in \S \ref{summary} we summarize the implications of these results
for the mass loss of this object.

\section{Observations and Data Reduction}\label{obs}

The observations of AFGL 5440 were made during an immersion occultation event on 28 Aug 2001 at approximately 02:44:00 UT.
The elevation of the Moon at the time of the event was 36\degree, and the Sun was 17\degree\ below the horizon.  The sky conditions
were not completely photometric but cloud cover was minimal and intermittent.  The position angle of the occultation event on
the lunar limb was 54\degree, and the lunar phase was 0.72.  We used the pMIRAS instrument on the McDonald Observatory 2.7-m
telescope.  The details of the instrument have been described by Harvey \& Wilson (2003), but we summarize the most
important characteristics here.  The instrument is essentially a long/wide slit, high-speed spectrophotometer using a
NaCl prism to disperse the light from the slit, which is then imaged onto a 32$\times$100 pixel portion of a 256$^2$
InSb array detector. The slit width is chosen to be the minimum acceptable in order to
minimize the background on the detector within the limitations imposed by the seeing conditions.  For these 
observations the slit width was set at about 5 arcsec for the
typical seeing of 1.5 arcsec.  The detector is read-out every 8 msec, with photon integration occuring over essentially the full 8 msec time.
Therefore, the Fresnel occultation fringe pattern is averaged over this 8 msec time (as well as over the 2.7-m telescope
aperture).  Because we use a refractive dispersive element, the dispersion/spectral resolution is not equal at all
wavelengths; the highest
dispersion is at the longest wavelengths, a feature that minimizes the background photon count at those wavelengths.  On average
over the 1 -- 4\micron\ waveband covered by the instrument, one pixel corresponds roughly to $\lambda/\Delta\lambda$ = 100. Because
of uncompensated seeing effects and the lower spectral dispersion at shorter wavelengths, the true resolving power at the
shortest wavelengths is R $\sim$ 20.
In the spatial direction the plate scale is 0.4 arcsec/pixel.
The instrument is read-noise limited shortward of 2.5\micron\ and background-limited longward of 3\micron.  A typical observation
consists of taking 5000 frames at a time roughly centered on the occultation event.  This is accomplished by using a buffer
that holds the most recent 5000 frames and terminating the data acquisition a few seconds after the event is observed on
a real-time display.  This is an important feature since the predicted times of occultation events are often in error by as much
as 10 seconds due to irregularities in the lunar limb as well as imperfect stellar astrometry.  
Our observations of the occultation event of the comparison star, IRC+00233, were obtained during an
immersion event on 29 Jun 2001, and those of HD 155292 were obtained during an immersion event on 26 Aug 2001.

The data reduction process consists of several typical steps.  Because the spectrum is being observed with much higher time
resolution than the seeing timescale, the spectrum moves around by several pixels over the course of an occultation event.
Therefore, to construct a light curve with minimal spectral blurring, the images must be shifted to correspond to the same
wavelength/pixel scale.  This is done with a simple cross-correlation algorithm that works well for the high S/N data
discussed here.  The spectrum is also not perfectly aligned with the X/Y axes of the detector, so we take out this tilt
as well during the processing to simplify later steps.  Because we use a ``long slit'', $\sim$ 10 arcsec, we can use the
sky measurements on either side of the stellar spectrum to provide an accurate and high-time-resolution sky subtraction.
For the data discussed here we produce a weighted average in the spectral direction that is two pixels wide and sum the pixels
in the spatial direction that have detectable signal.  We have experimented with more elaborate photometric extraction
schemes, but this technique appears to produce S/N ratios as high as any more complex algorithms.   We have also experimented
with various flat-fielding methods but have found only a small improvement in S/N with these methods.  The end result of
the data reduction process
is a sequence of $\sim$ 100 light curves for which we extract a few hundred frames centered on the occultation event.
The frames that are taken well before the Fresnel fringe pattern of the occultation becomes evident are used to estimate the noise
level in the data, due both to read-noise, background photon statistics, and the often non-negligible amount of
seeing/scintillation noise caused by the atmosphere.  

Because the exact location of the spectrum on the detector
varies both due to seeing and between observing runs after adjustments to the instrument, we perform the wavelength
calibration by fitting the NaCl dispersion function to the observed positions of the J, H, K, and L atmospheric
transmission maxima in the actual data for each occultation event.  The accuracy of this calibration is probably
good to $\pm$ .01 $\lambda$ throughout the 1 -- 4\micron\ region that is observed.
Because of variable instrumental efficiency depending on the placement of the star image on the input slit, and
the common occurrence of non-photometric sky conditions during occultation events, we specifically do not
attempt to derive a flux calibration for our data.  We have, however, compared the relative signal from AFGL 5440
to that of relatively well characterized stars observed on the same night and conclude that the stellar magnitudes
in the published literature for AFGL 5440 are consistent with values that we would have derived from
our signal strengths to within $\pm$ 30\%.

\section{Source Modeling}\label{model}

The fringe pattern of a lunar occultation event is a convolution of the pattern for a point source over several parameters that all act
to blur the fringes.  These parameters include: the telescope aperture, the integration time of the detector, the
wavelength bandwidth of the observation, and, most importantly, the source size/structure.  In order to extract the
source size or more detailed properties of the spatial structure, the typical procedure is to model the combination of
all the above ``blurring'' parameters with various possible source models to find the best fit to the data (e.g.
Nather \& McCants 1970; Richichi et al. 1995).
An additional uncertainty in the observations is the basic frequency of the fringe pattern, i.e. the speed of the
lunar shadow.  Although this parameter is calculated by the software that we use to predict occultation events, small uncertainties
in the shape of the lunar limb (roughness due to craters, etc) can produce differences in the predicted shadow speed up
to several tens of percent.  Therefore, this parameter must also be fit in addition to the parameters that blur the fringes.

We began our modeling process by assuming a uniform disk 
as the simplest possible model with which to try fitting the data for the 
observed objects, AFGL 5440, IRC+00233, and HD 155292.
We use a simple $\chi^2$ test for the best model, allowing one or more parameters to vary during the process.
Typically we first allow both the size and lunar shadow velocity to vary until we find an approximate fit to
both.  This fit can be done either individually at each wavelength or globally using the entire waveband.

\subsection{Comparison Stars}

For IRC+00233 and HD 155292 we assumed that a single source size was likely to be appropriate 
for the entire wavelength range
within our observational uncertainties,
and we fit the observations globally for source size and lunar velocity.  
Rough estimates of the angular sizes of these two stars can be derived simply by assuming that they are
blackbodies of the effective temperatures given by their spectral types.  Using the empirical relation between
angular size and B-K color from van Belle (1999) for IRC+00233 (B = 11.06, K = 1.95),
we would expect an angular size 
of 3.5 mas; the star is, however, likely to be mildly variable, so the size at the time of our
observations might have been different by $\pm$ 20\%.  For HD155292 (B = 11.0, K = 4.9) a similar
calculation gives
an angular size of 0.6 mas.   Since typical departures from a uniform
disk model are at the level of 10 -- 30\% (e.g. Thompson, Creech-Eakland \& van Belle 2003; and Scholz 2003), 
the resolution required to detect them reliably for even IRC+00233 is below 1 mas.  
Based on tests we have done with model data,
this is beyond the capabilities of our current data set which is limited by both
spectral resolution and signal-to-noise ratio to accuracies on the order of $\pm$2 mas. 

\begin{figure}
\plotfiddle{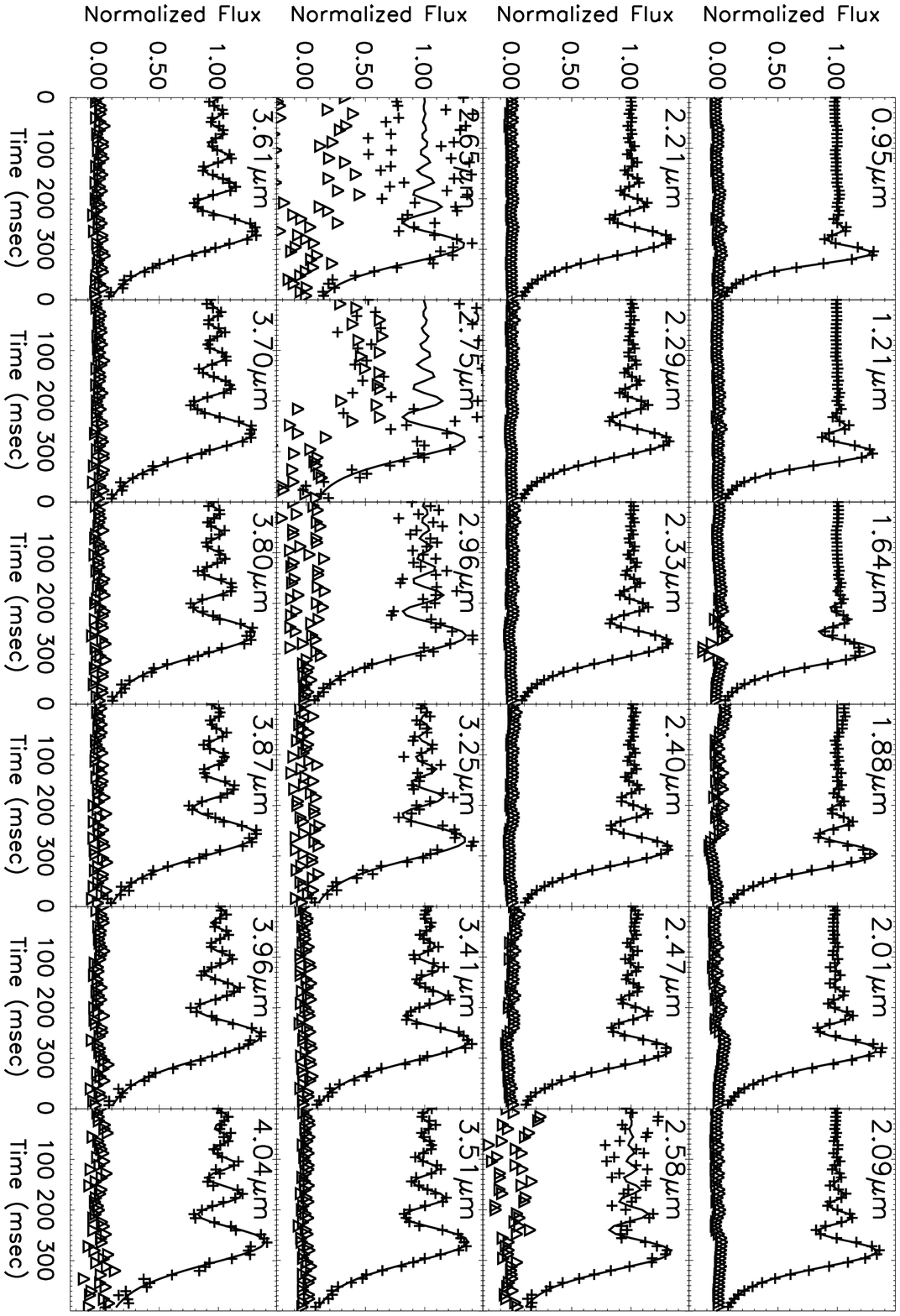}{8.0in}{180}{75}{75}{220}{590}
\figcaption{\label{hatlams} Plots of observed light curves (crosses) versus computed model light curves for the
4.5 mas uniform disk model of IRC+00233 at a sampling of the range of wavelengths observed between 1 and 4\micron.
In each panel the lower curve of triangles shows the difference of observed minus model.}
\end{figure}

\begin{figure}
\plotfiddle{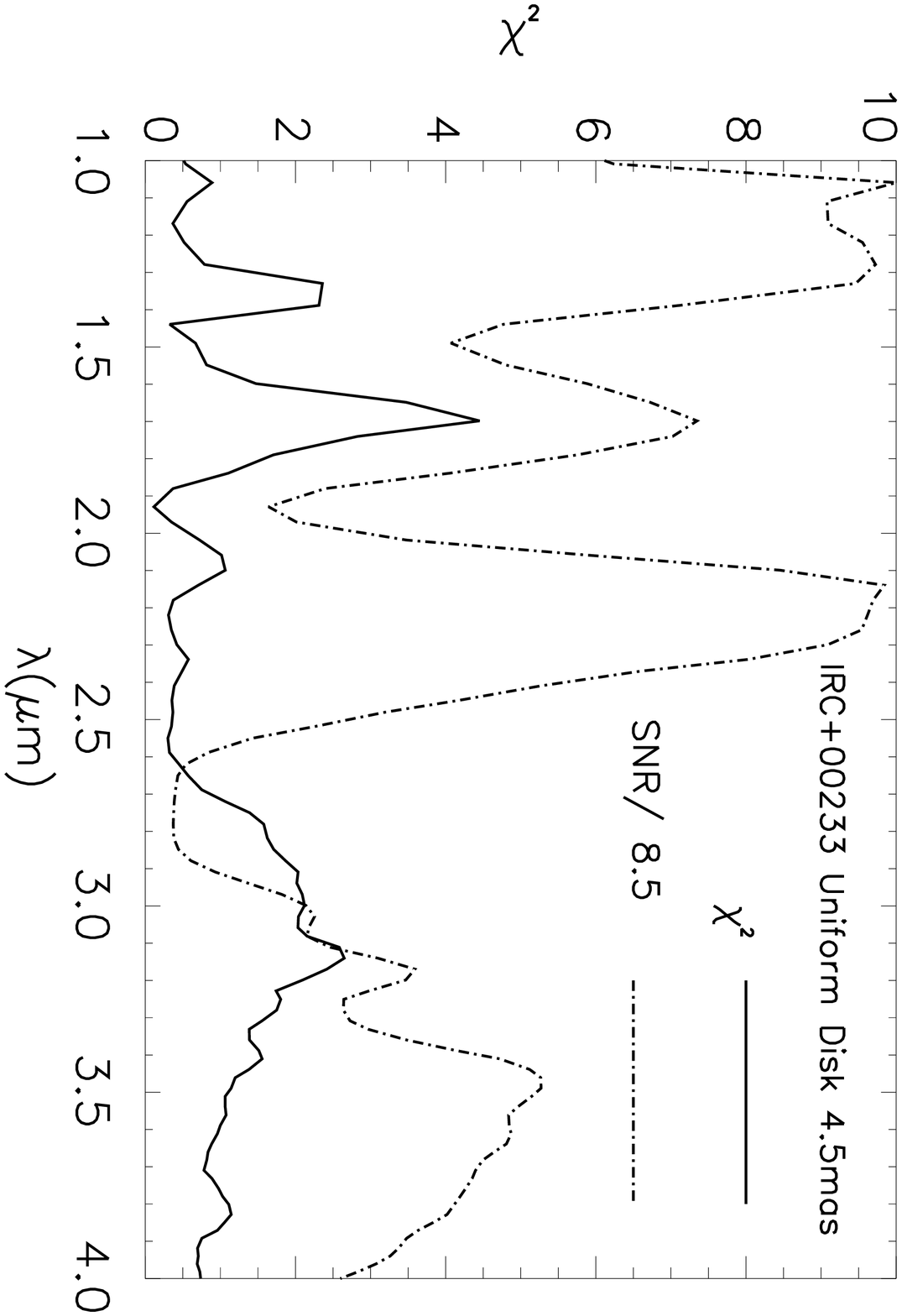}{5.0in}{90}{65}{65}{270}{0}
\figcaption{\label{hatchi} Plots of the $\chi^2$ (solid) for the model fit of the 4.5 mas uniform disk to the observed
light curve for IRC+00233 over the range of wavelengths observed between 1 and 4\micron.  The observed signal-to-noise
ratio is also shown (dashed) as a function of wavelength for comparison.}
\end{figure}

\begin{figure}
\plotfiddle{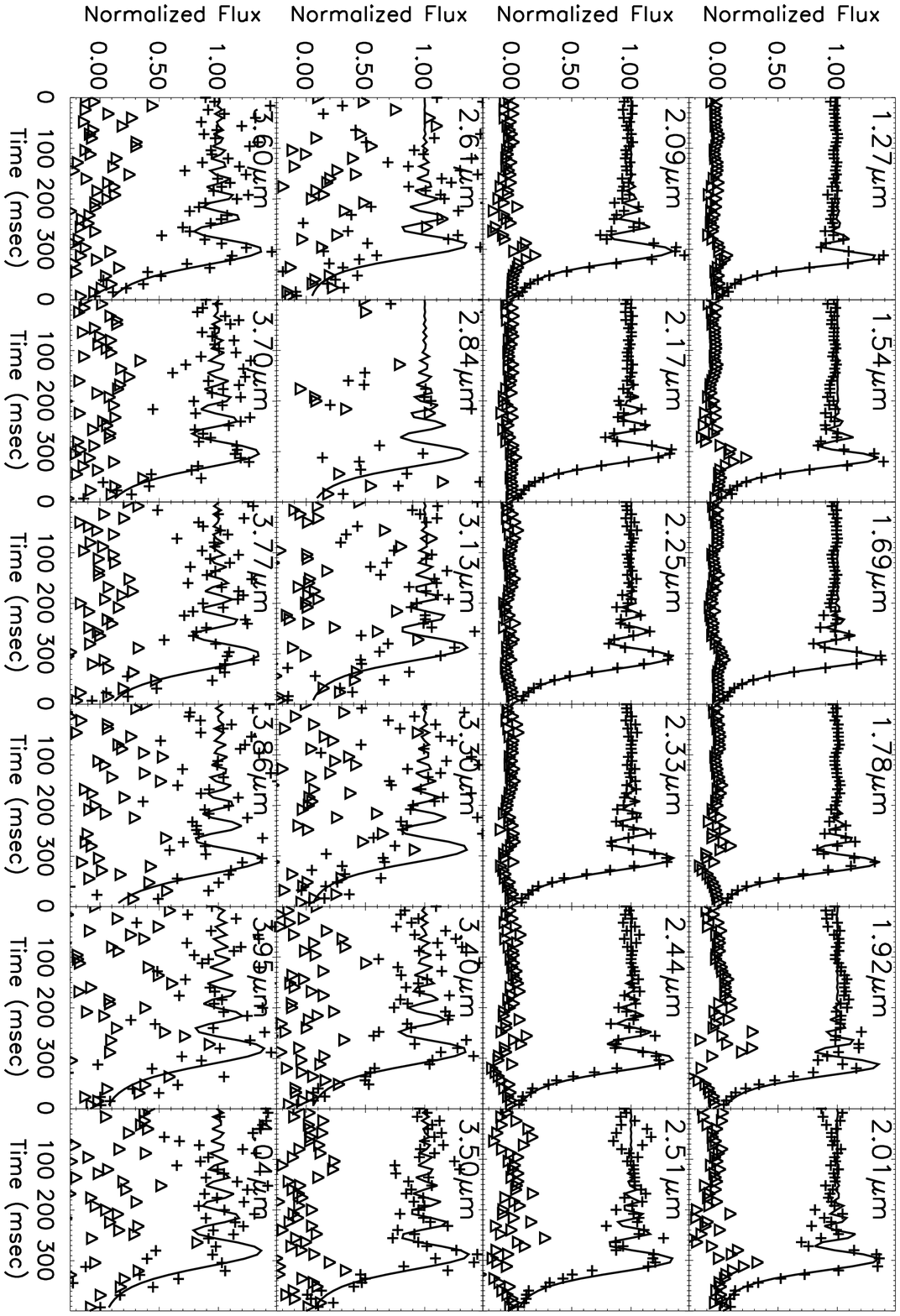}{8.0in}{180}{75}{75}{220}{590}
\figcaption{\label{hatlamshd} Plots of observed light curves (crosses) versus computed model light curves for the
2.0 mas uniform disk model of HD155292 at a sampling of the range of wavelengths observed between 1 and 4\micron.
In each panel the lower curve of triangles shows the difference of observed minus model.}
\end{figure}

\begin{figure}
\plotfiddle{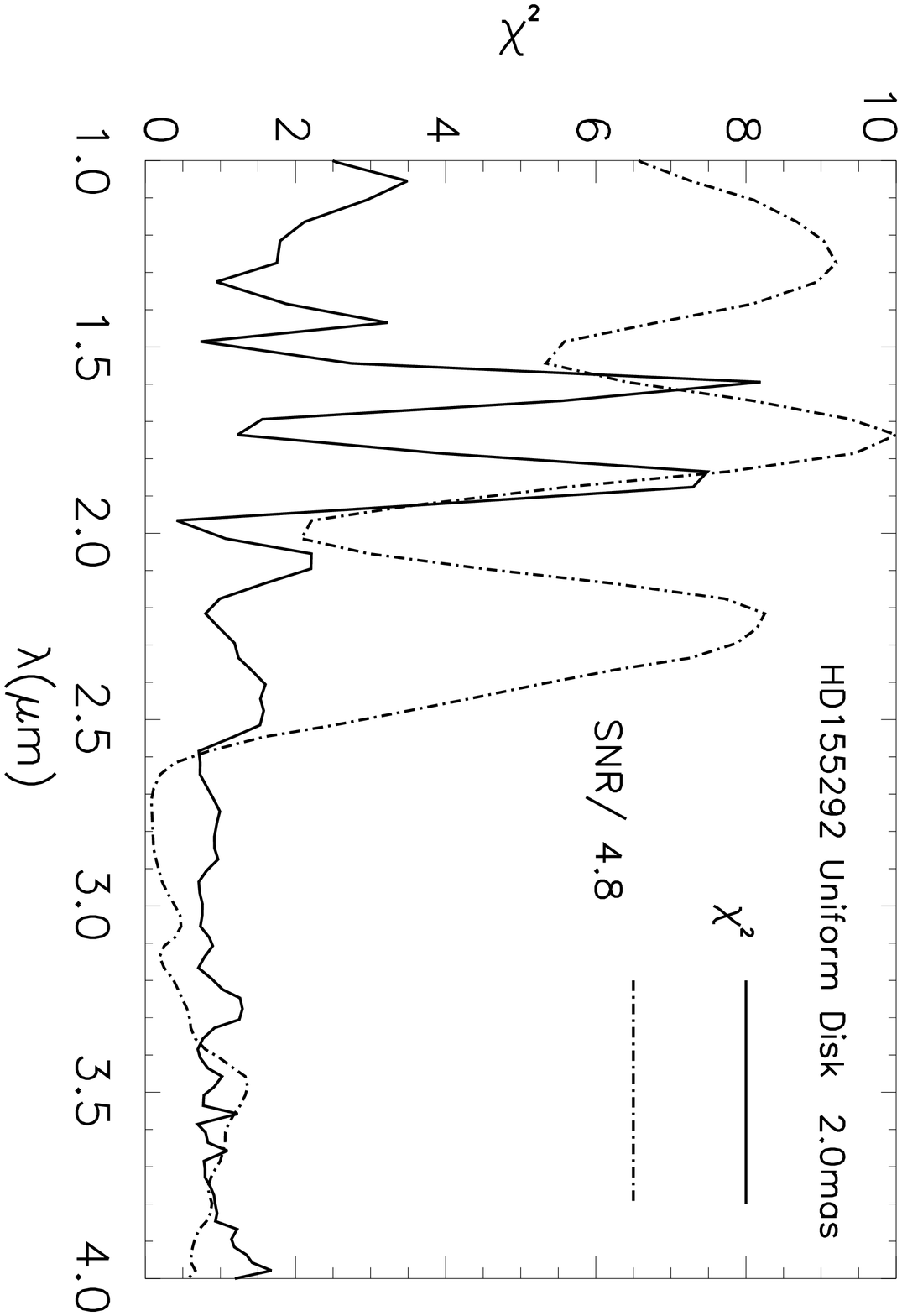}{5.0in}{90}{65}{65}{270}{0}
\figcaption{\label{hatchihd} Plots of the $\chi^2$ (solid) for the model fit of the 2.0 mas uniform disk to the observed
light curve for HD155292 over the range of wavelengths observed between 1 and 4\micron.  The observed signal-to-noise
ratio is also shown (dashed) as a function of wavelength for comparison.}
\end{figure}

The best fit size for IRC+00233
is between 4.5 and 5.0 mas.  Figure \ref{hatlams} shows observed and model light
curves for a model assuming a 4.5 mas uniform disk for a subset of the wavelengths observed.  Figure \ref{hatchi} 
shows the $\chi^2$ and signal-to-noise ratio as
a function of wavelength over the entire observed band for this model.  Both figures show that this model provides a very
good fit to the observations except for a small range of wavelengths around $\sim$ 1.7\micron\ where a substantially
larger size would provide a better fit.  We do not have a good explanation for this discrepancy; it may be due to
some systematic noise effect or to a real difference in the stellar photosphere in that region.  
For comparison
Schmidtke et al. (1986) observed this same star in an occultation event using narrow-band
filters near 1.6 and 2.2\micron.  They found a size at those wavelengths of $\sim$ 3 mas, similar to but smaller
than our calculated blackbody
size.  
For HD 155292 all uniform-disk models with a size less than about 3 mas were able to fit the data reasonably well
(Figures \ref{hatlamshd} and \ref{hatchihd}).
Since the blackbody size of the star is less than a milli-arcsecond, this is consistent with the expected
uncertainties in our data and modeling.  The signal-to-noise ratio for this star was low enough that
we had no effective narrow-band information beyond 2.5\micron, and at the shorter wavelengths some
periodic electronic pickup had a non-negligible effect on the observed fringe patterns as well.
Note that for both these comparison stars there are wavelengths with reasonable S/N as shown in 
Figures \ref{hatlams} and \ref{hatlamshd} where there is a less than adequate fit, so these are
difficult to explain solely as due to telluric atmospheric absorption effects. 

\subsection{AFGL 5440}

\begin{figure}
\plotfiddle{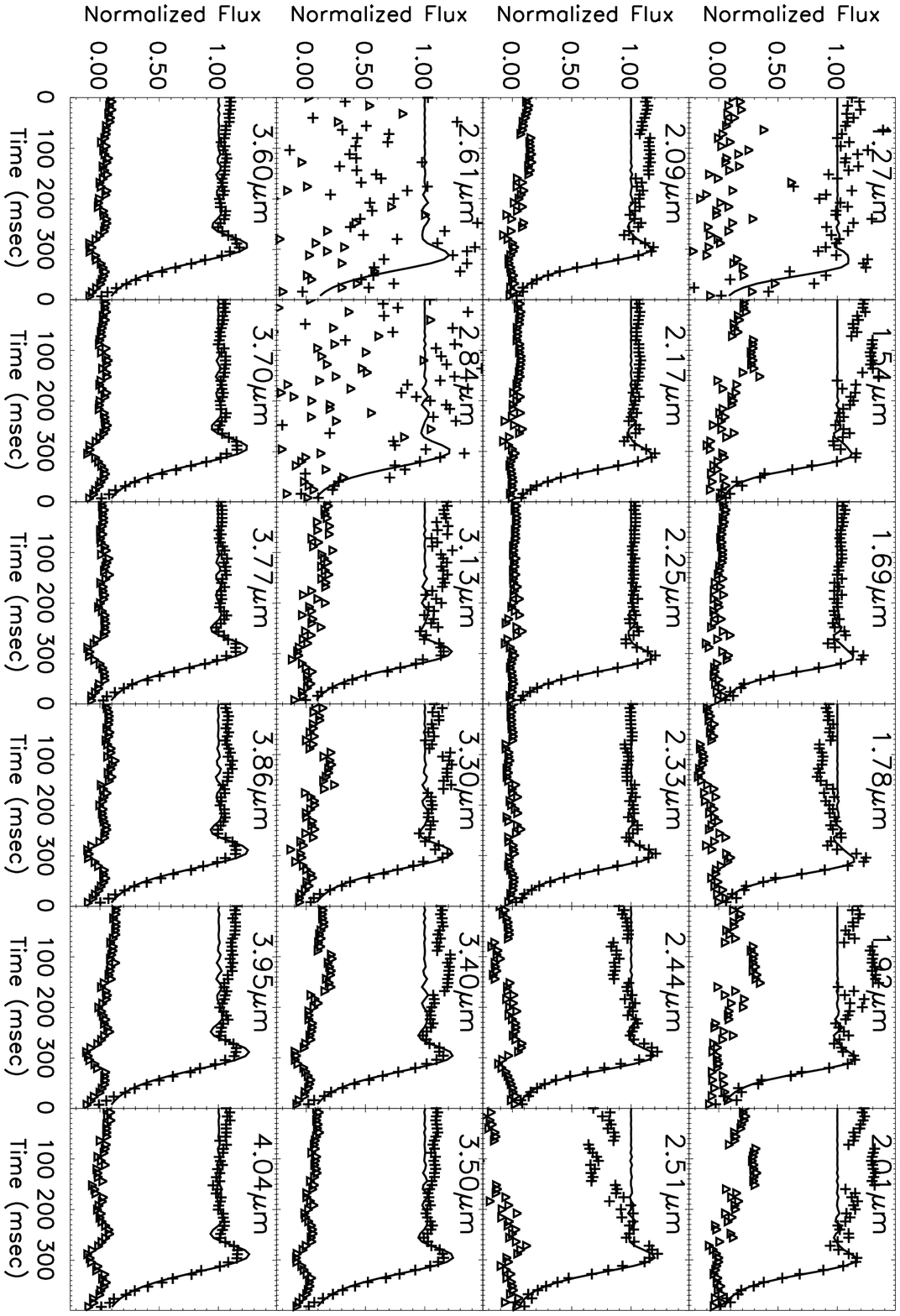}{8.0in}{180}{75}{75}{220}{590}
\figcaption{\label{hat11lams} Plots of observed light curves (crosses) versus computed model light curves for the
11 mas uniform disk model for AFGL 5440 at a sampling of the range of wavelengths observed between 1 and 4\micron.
In each panel the lower curve of triangles shows the difference of observed minus model.}
\end{figure}

\begin{figure}
\plotfiddle{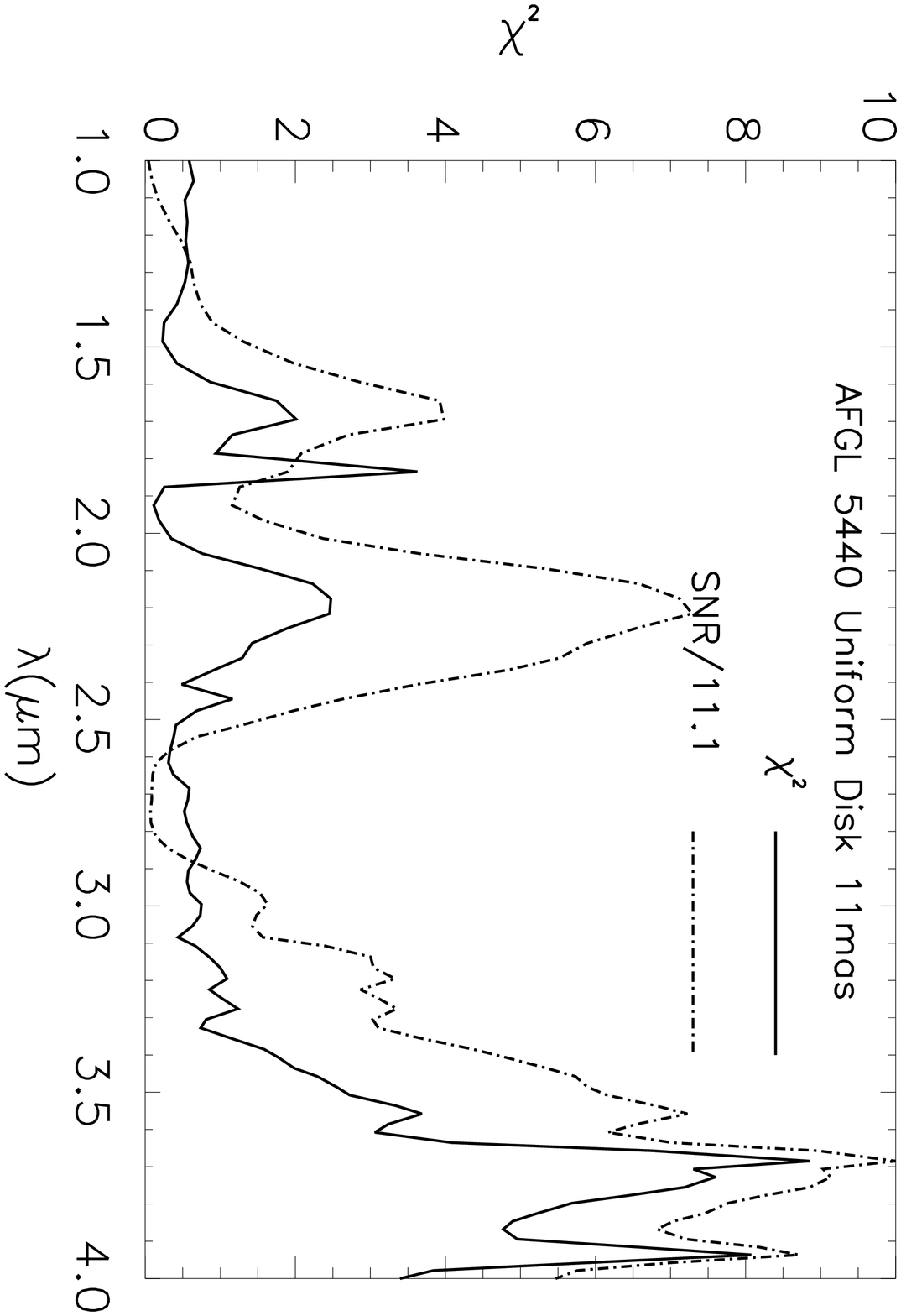}{5.0in}{90}{65}{65}{240}{0}
\figcaption{\label{hat11chi} Plots of the $\chi^2$ (solid) for the model fit of the 11 mas uniform disk to the observed
light curve for AFGL 5440 over the range of wavelengths observed between 1 and 4\micron.  The observed signal-to-noise
ratio is also shown (dashed) as a function of wavelength for comparison.}
\end{figure}

For AFGL 5440 a quick glance at the
observed light curves (Figure \ref{hat11lams}) indicated that a single source size was unlikely to fit over the entire 1 -- 4\micron\ bandwidth
(Harvey \& Wilson 2003).  This suggests that we are seeing the combination of the emission from the central star and
a circumstellar shell of material due to mass loss from the star.  To demonstrate the poor fit with a single size
uniform disk, we show in Figures \ref{hat11lams} and \ref{hat11chi} the results of trying to fit the data with one example uniform disk,
11 mas.  As can be seen in the plots of $\chi^2$ as well as the observed versus model light curves, the
11 mas disk provides a passable fit in the mid-range of wavelengths, 2.5 -- 3.2\micron, but produces
fringes that are too sharp at the longer wavelengths and too broadened at the shortest wavelengths. 

This result motivated us to pursue a full radiative transfer model for the object that could be used to compare both the
size constraints provided by our occultation data {\it and} the spectral energy distribution \citep{guand06} which contains
important and different information about the relative amount of dust at different temperatures.  The DUSTY
code \citep{ivez99} was originally created, in fact, for modeling the emission from AGB stars surrounded by mass-loss shells.  Its
output includes model source images as well as the total energy distribution, so it is ideal for our purposes.
Our approach to using the code was to choose a particular combination of input parameters and then vary the
dust optical depth to find the best fit to the energy distribution for those parameters.  We then used the
output source images that were computed as a function of wavelength to calculate expected occultation fringes for the model and
compared those to the observed fringes.  Since AFGL 5440 has been classified as a carbon-rich star,
we assumed a carbon-rich dust composition, typically some combination of amorphous carbon, silicon carbide,
and graphite as the major constituents.
The other critical parameters for the models are the dust temperature
at the inner radius and the radial density gradient.  The outer radius of the dust shell makes essentially no difference
to the observed characteristics at $\lambda <$ 60\micron\ as long as it is at least 100 times the inner radius.
The goal of our modeling was to find some reasonable fit to our occultation data and the rough spectral
energy distribution for the circumstellar dust shell; we have not attempted to extract
details of the stellar photosphere or attempt a thorough examination of all possible dust size/composition
models since our data do not bear directly on those issues for reasons of wavelength coverage and
spectral resolution.  In particular, we did not try to find any better than superficial agreement with
the IRAS LRS data.

\begin{figure}
\plotfiddle{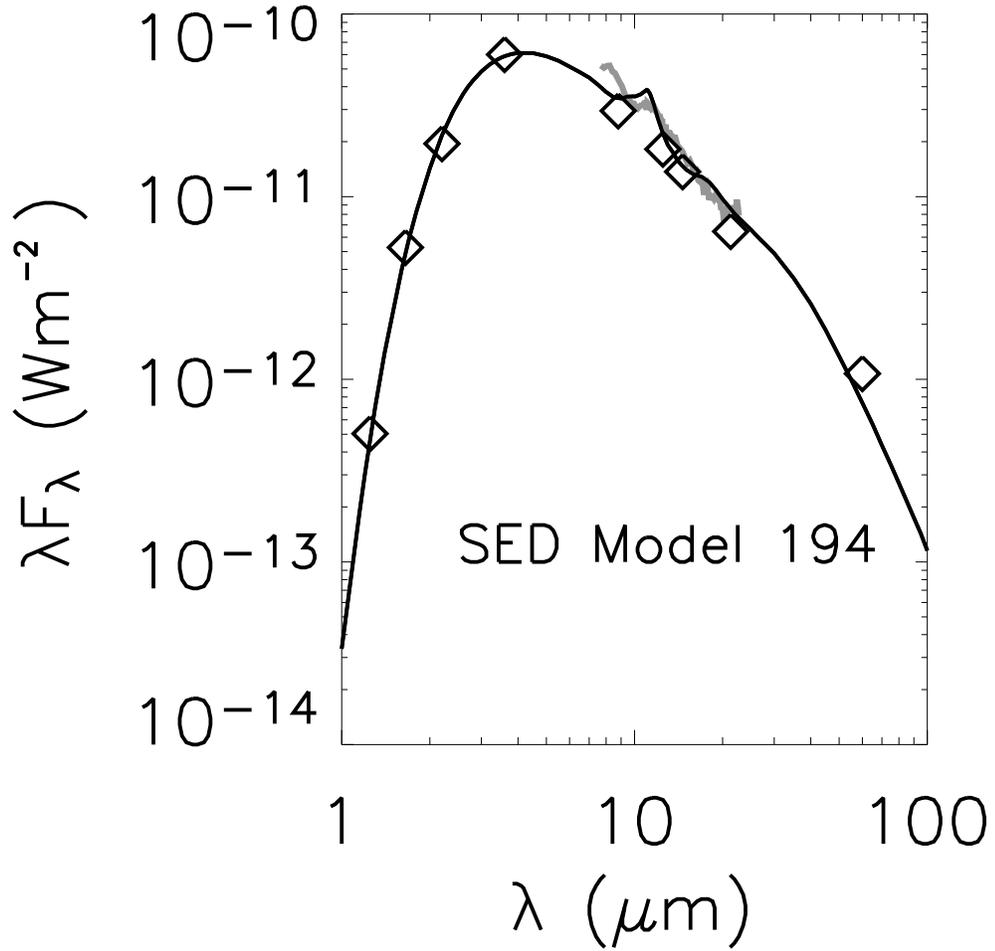}{5.0in}{0}{100}{100}{-310}{-100}
\figcaption{\label{sedbest} Plot of the observed spectral energy distribution of AFGL 5440 versus that predicted
by model 194, the best fit model.  The open diamonds show the values from \citet{guand06} and \citet{gug93}.  The light
grey line indicates the IRAS LRS spectrum.}
\end{figure}

\begin{figure}
\plotfiddle{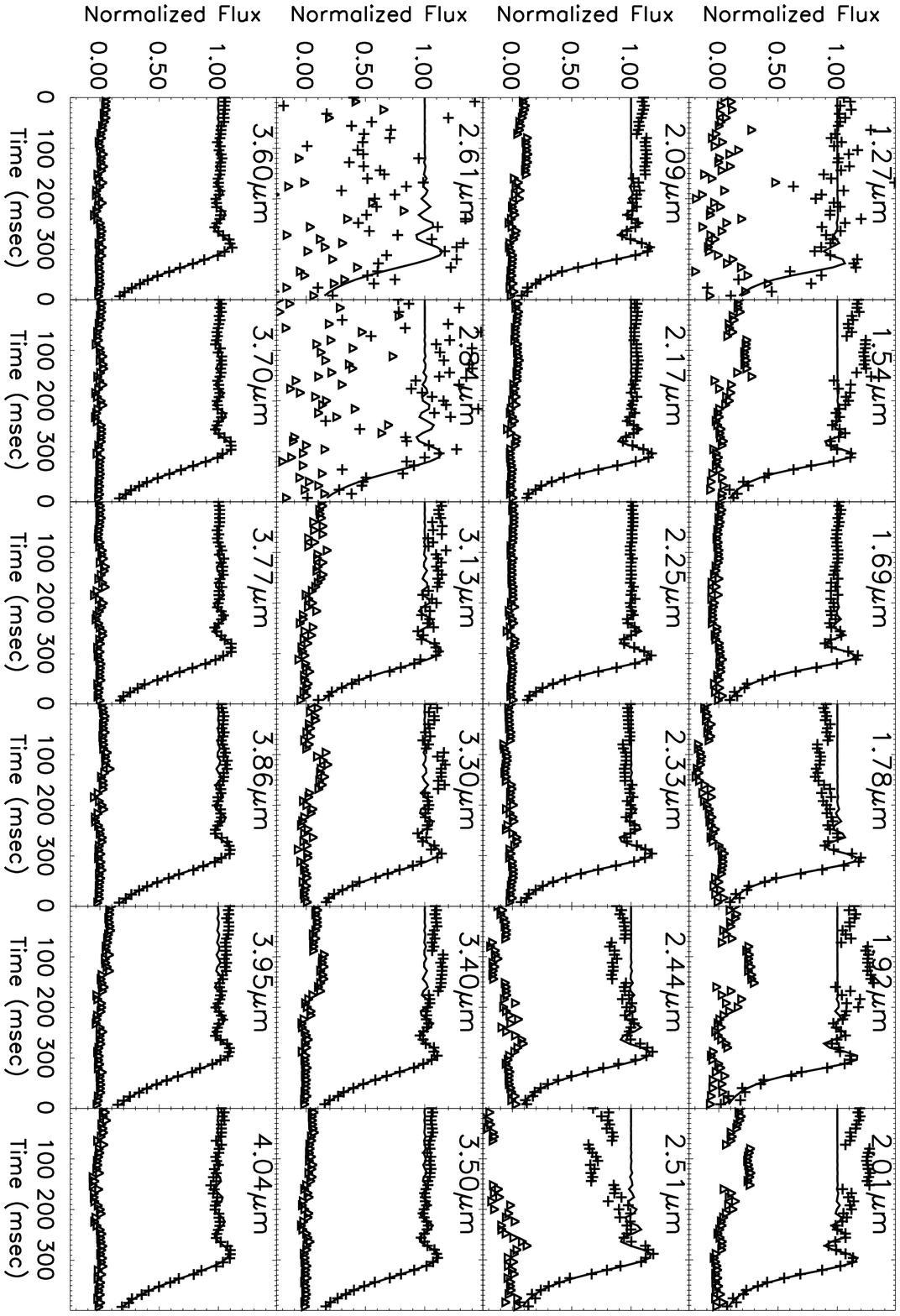}{8.0in}{180}{75}{75}{220}{590}
\figcaption{\label{bestlams} Plots of observed light curves (crosses) versus computed model light curves for model 194
for AFGL 5440 at a sampling of the range of wavelengths observed between 1 and 4\micron.
In each panel the lower curve of triangles shows the difference of observed minus model.}
\end{figure}

\begin{figure}
\plotfiddle{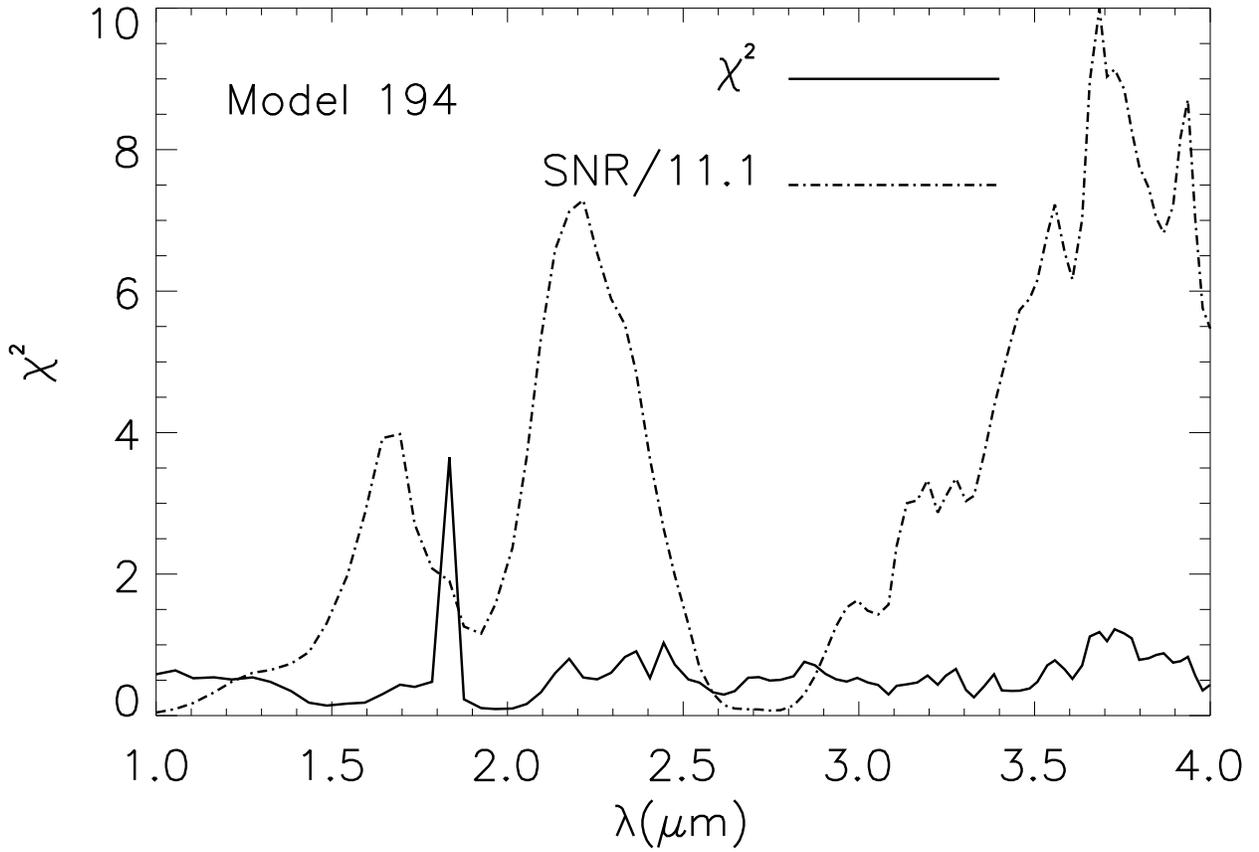}{5.0in}{90}{65}{65}{260}{0}
\figcaption{\label{bestchi} Plots of the $\chi^2$ (solid) for the fit of model 194 (the best fit) to the observed
light curve for AFGL 5440 over the range of wavelengths observed between 1 and 4\micron.  The observed signal-to-noise
ratio is also shown (dashed) as a function of wavelength for comparison.}
\end{figure}

\begin{figure}
\plotfiddle{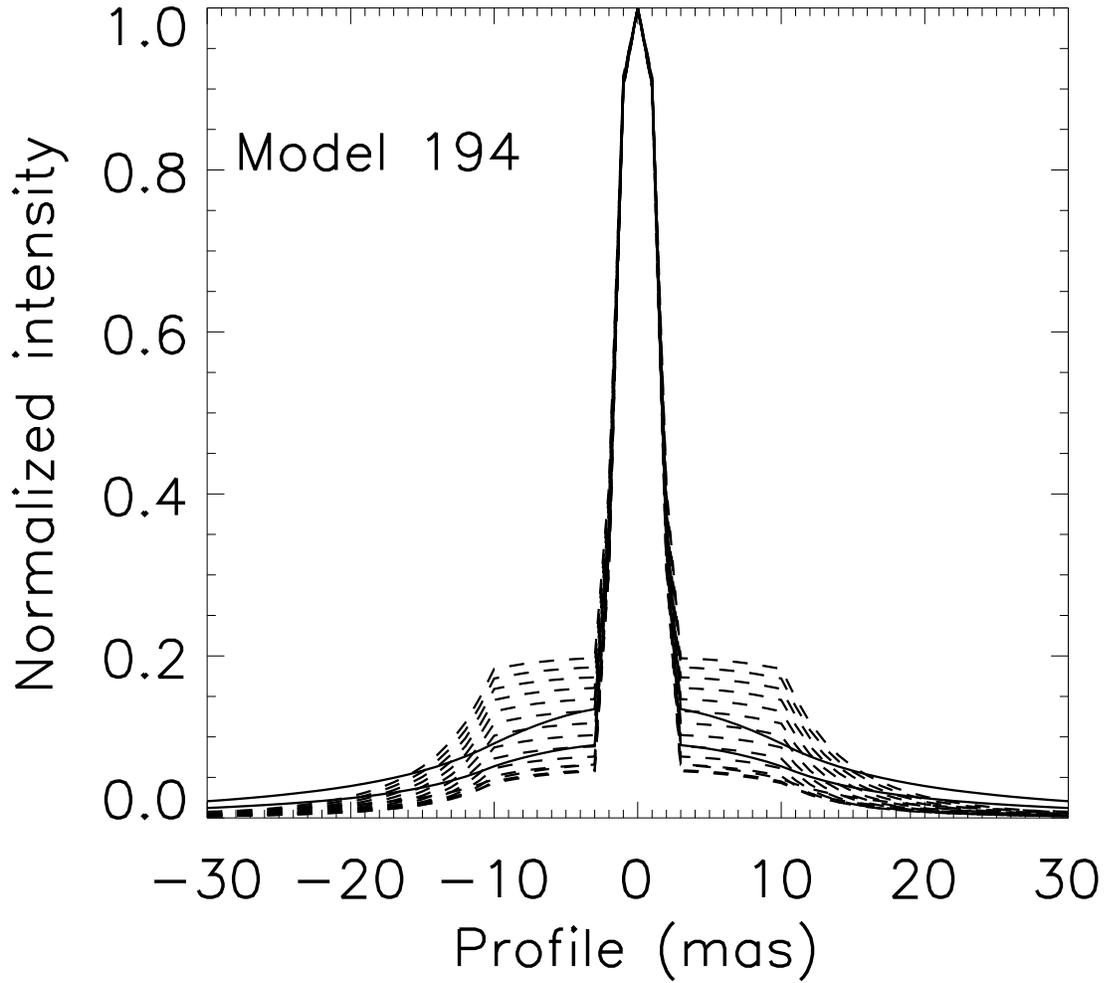}{5.0in}{0}{100}{100}{-330}{-90}
\figcaption{\label{liprobest} Plots of the model spatial distributions for AFGL 5440 predicted
by model 194, the best fit model.  The curves are drawn for 1.0 and 1.2\micron\ (solid, 1.0\micron\ is the more extended),
and 1.4, 1.6, 1.8, 2.0, 2.2, 2.4,
2.6, 2.8, 3.0, 3.2, 3.4, 3.6, 3.8, and 4.0\micron, with the longest wavelengths being the most extended of these.}
\end{figure}

\begin{figure}
\plotfiddle{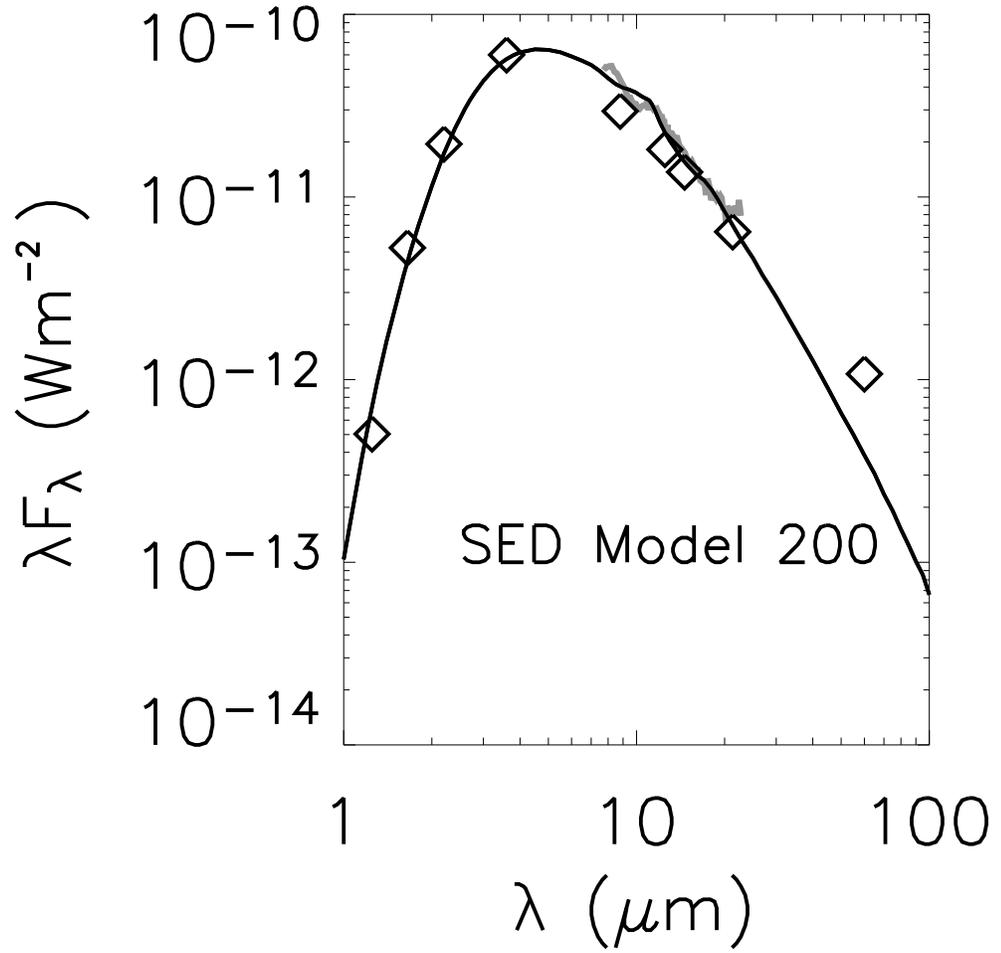}{5.0in}{0}{100}{100}{-310}{-100}
\figcaption{\label{sedbestamorph} Plot of the observed spectral energy distribution of AFGL 5440 versus that predicted
by model 200, the best fit model that does not use graphite as for fig. \ref{sedbest}.}
\end{figure}

\begin{figure}
\plotfiddle{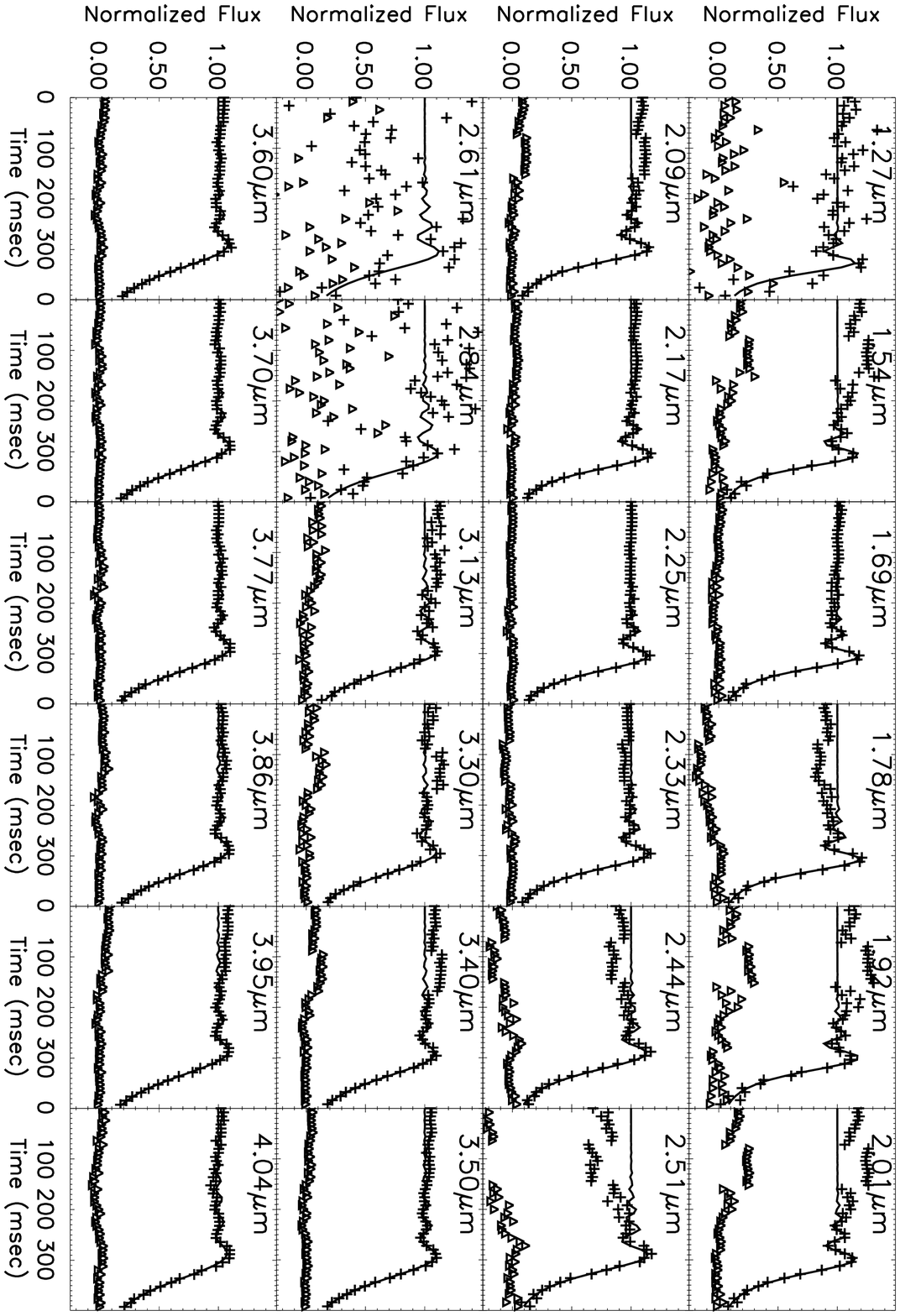}{8.0in}{180}{75}{75}{220}{590}
\figcaption{\label{bestlamsamorph} Plots of observed light curves (crosses) versus computed model light curves for model 200
for AFGL 5440 at a sampling of the range of wavelengths observed between 1 and 4\micron.
In each panel the lower curve of triangles shows the difference of observed minus model.}
\end{figure}

\begin{figure}
\plotfiddle{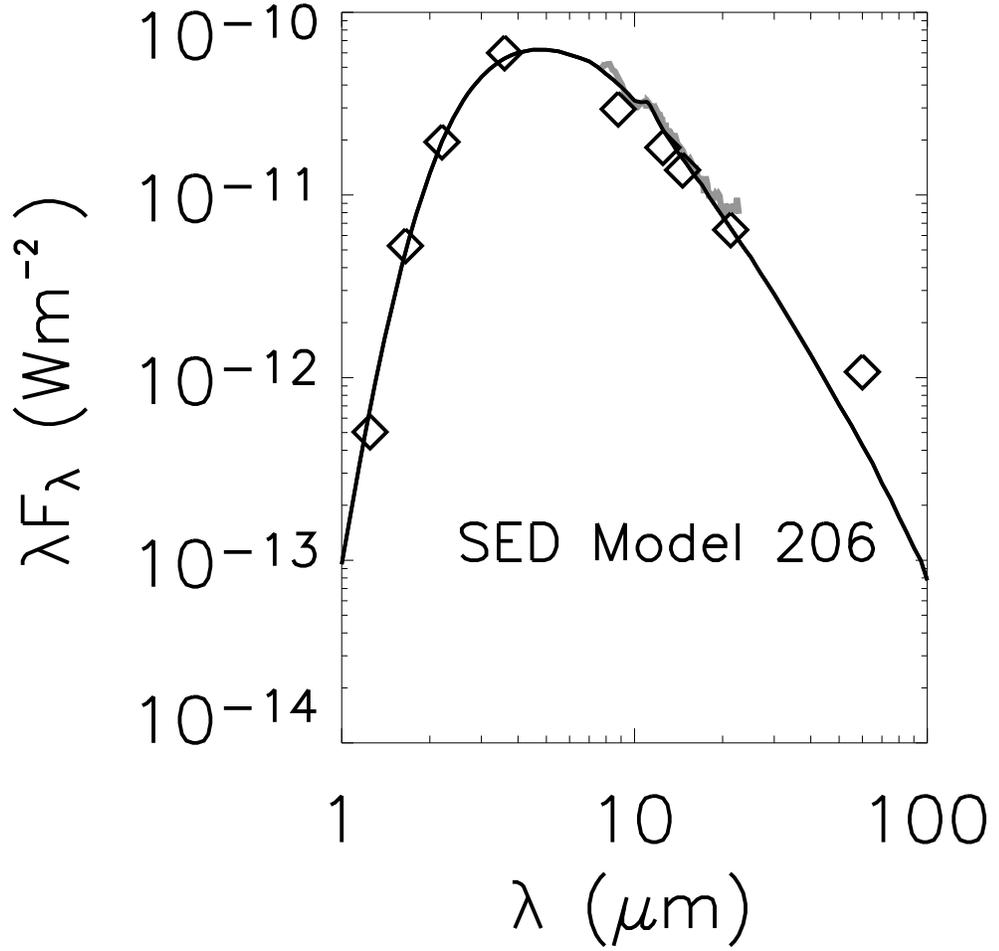}{5.0in}{0}{100}{100}{-310}{-100}
\figcaption{\label{sedbad} Plot of the observed spectral energy distribution of AFGL 5440 versus that predicted
by model 206, illustrating the fit for a shell with a cooler inner dust temperature as for fig. \ref{sedbest}.}
\end{figure}

\begin{figure}
\plotfiddle{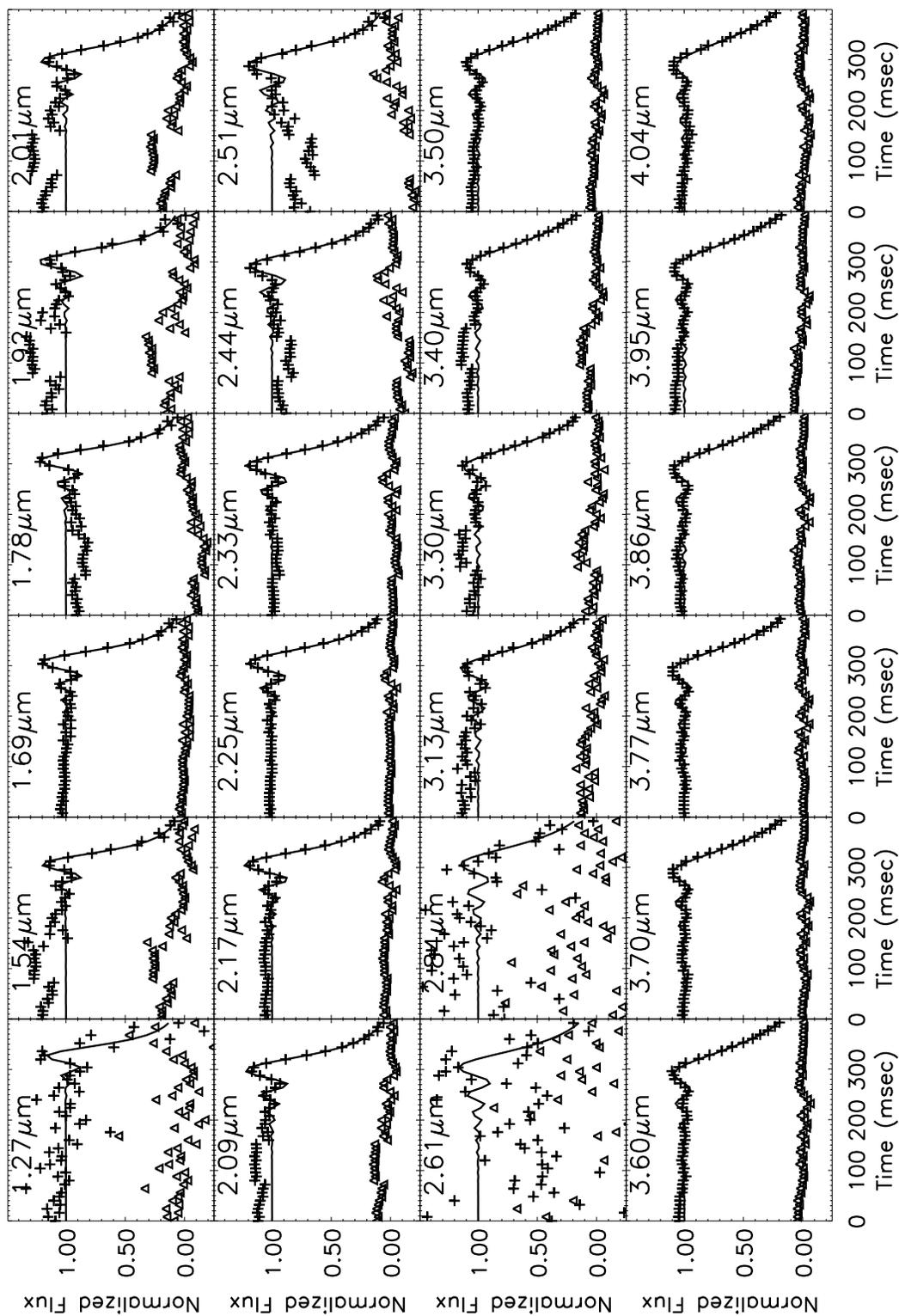}{8.0in}{180}{75}{75}{220}{590}
\figcaption{\label{badlams} Plots of observed light curves (crosses) versus computed model light curves for model 206
for AFGL 5440 at a sampling of the range of wavelengths observed between 1 and 4\micron.  
In each panel the lower curve of triangles shows the difference of observed minus model.
Note the larger residuals
at the longer wavelengths than for the best fit model 194.}
\end{figure}

We explored more than 200 models to understand the effect of varying the input parameters on the quality of
the fit.  Basically all the models that provided an approximate fit to the occultation observations and
the spectral energy distribution had several features in common.  First, the dust temperature at the inner
radius of the dust shell was of order 950K $\pm$ 50K.  Models with a maximum dust temperature below 900 K did
not have enough hot dust to fit the occultation fringes at the longer wavelengths, while models with hotter inner
edges had even more difficulty than the best-fit model in reproducing the energy distribution longward of 10\micron.  
Secondly, the radial density gradient of models with a reasonable fit was close to
r$^{-2}$ (or to DUSTY's calculation of the gradient appropriate for a radiatively driven wind
which approximates an r$^{-2}$ distribution for large radii).  
Models with a density gradient of r$^{-1.8}$ came closer to producing the IRAS 60\micron\ flux, but did not
fit the shape of the 5 -- 20\micron\ energy distribution well.  Models with a density gradient of r$^{-2.2}$
can fit the energy distribution out to 20\micron\ and also provide a good fit to the occultation results, but
have substantially worse fits to the IRAS 60\micron\ flux than our best fit models.
The optical depth for best fit at the fiducial wavelength
of 5\micron\ was typically in the range 0.3 -- 0.6.  Finally, the dust composition that provided the best fit included
a small amount of SiC together with comparable amounts of amorphous carbon and graphite.  Other carbon-rich
compositions produced reasonable fits except at the longest wavelengths or in the 11\micron\ SiC feature.
Note that there is a range of optical properties for different forms of amorphous carbon \citep{ander99} that
we did not explore.
We experimented with two grain size distributions, the MRN slope \citep{mathis77}, and another, the KMH shape \citep{kim94}, used by
\citet{ivez96} for models of IRC+10216, that has a smoother fall-off on each end.
We found that we could obtain reasonably good fits with either distribution.
Figures \ref{sedbest} and \ref{bestlams} show the fits to the spectral energy distribution and to the occultation light curves for the
best-fit model.  Figure \ref{bestchi} shows the quality of the fit versus wavelength, and figure \ref{liprobest} illustrates
the spatial profiles of this best-fit model.   Figures \ref{sedbestamorph} and \ref{bestlamsamorph}
 likewise show for comparison the fits for a model discussed below
that does not use any graphite.  Finally, figures \ref{sedbad} and \ref{badlams}
show results for a third model that fits the energy distribution but
gives a poor fit to the occultation results because of a lack of enough warm dust close to the star.

\section{Discussion and Summary}\label{summary}

Model dust shells have been computed for a number of carbon-rich AGB stars by various authors.
Le Bertre, Gougeon, \& Le Sidaner (1995) and Le Bertre (1997) found that the energy distributions
for nearly two dozen carbon-rich stars could be modelled with very similar dust shell parameters.
They found a common value of maximum dust temperature of order 950 K with shell density gradient of
$\rho \propto r^{-2}$ for spherically symmetric shells.  The best fit dust property implied a dust
emissivity, $\epsilon \propto \lambda^{-1.3}$, consistent with that expected for an amorphous carbon-rich dust
composition as also found by \citet{jura04}.  These conclusions were enhanced by their ability to fit the energy distributions over
a range of phases of the observed variability of many of the stars.  On the other hand, Suh (1997)
suggested that a ``superwind'' phase of mass loss could improve model fits to the energy distributions
for a number of carbon-rich AGB dust shells by enhancing the emission shortward of 30\micron\ because
of a high rate of mass loss in recent times, e.g. at small radii where most of the emission would
be due to hotter dust.  \citet{ivez96} used a self-consistent radiatively driven hydrodynamic model for
the density distribution around IRC+10216 and were able to fit both the spectral energy distribution
and the near-infrared angular visibility data from speckle observations.  Interestingly they derived
an inner dust shell temperature substantially lower than ours and most other studies, $\sim$ 750K.
\citet{winters97} computed
a full time-dependent model of periodic outflow from AFGL 3068 and were able to produce a good
fit to the energy distribution and observed light curves.  Virtually all of these
modeling efforts used dust emissivities appropriate for some form of amorphous carbon (usually
including SiC), but with no graphite (see also Lorenz-Martins 2001), contrary to our best model fit above.  The model of
\citet{ivez96} utilized a grain size distribution that included substantially larger grains than most
earlier models.

Our lunar occultation observations add constraints to these results most directly in defining the 
amount of dust in
the innermost region around AFGL 5440, since our longest observed wavelength is 4\micron.  Basically, our data
imply that the density of grains emitting in the 3 -- 4\micron\ spectral region must be sufficiently
large to produce the substantial fringe ``blurring'' observed in our light curves.  Graphically,
this amount of emission is illustrated in figure \ref{liprobest} showing the source profiles for the best-fit
model.  For any given assumed dust emissivity, this constraint then implies a fairly narrow
range of optical depth and ratio of near-infrared to mid-infrared dust emission.  Thus, we
constrain the dust density gradient in the innermost part of the circumstellar shell.
Finally, as mentioned above, the fact that we were unable to find a satisfactory fit to
the data with {\it any} model having a maximum dust temperature lower than 900 K is a strong
constraint on the location of the inner edge of the dust shell.  This result follows
from the fact that dust cooler than 900 K cannot emit sufficiently to produce the required 
amount of 3 -- 4\micron\ emission.  Note that this temperature is nearly a factor of two below the
expected condensation temperatures for dust around these stars \citep{egan95}.  The physical value of this inner radius for AFGL 5440
for the dust properties of the best-fit model is 3.7$\times 10^{14}$ cm, or an angular radius
at the assumed distance of 2.25 kpc of 11 mas.  For reference the angular {\it diameter} of the central star is
of order 3 mas, based on a blackbody approximation for its likely photospheric temperature 
of 2500K $\pm$ 300K and 2\micron\ magnitude.

Since our modeling is the first of which we are aware for this particular star over the entire infrared wavelength
region, we also discuss here the constraints on the circumstellar 
cloud properties implied
by the overall energy distribution.  
As mentioned above, our best-fit model utilized roughly
equal amounts of amorphous carbon and graphite in addition to the small amount of SiC required
to fit the 11\micron\ feature seen in the IRAS LRS spectrum.  This result is contrary to the
large amount of evidence against substantial amounts of graphite in stars like AFGL 5440.  The
factor that drove us to include graphite was its relatively flat emissivity vs. wavelength
dependence between 10 and 50\micron\ that enabled us to fit the IRAS 60\micron\ flux.
Models with only amorphous carbon and SiC failed to fit that flux by factors of 3 or more.
Figures \ref{sedbestamorph} and \ref{bestlamsamorph} discussed above show the energy distribution and model 
light curves for the best-fitting model
that uses only amorphous carbon and SiC, and which uses the radiatively driven hydrodynamic density gradient of
Ivezic \& Elitzur (1996) as computed by DUSTY.  The light curves fit the observed occultation
data nearly as well as those for the best-fit model.  

Clearly, however, the IRAS 60\micron\ flux cannot
be fit by any similar model without graphite, and substantial modifications would have to be made to the
assumed density law in the outer regions or to the dust emissivity in order to come close
to fitting the 60\micron\ flux.  Explaining this problem is beyond the scope of our
study, but the fact that there is abundant evidence for non-constant outflow from carbon
stars \citep{fong03,mauron00} provides a convenient (if ad hoc) explanation.  If the mass-loss rate were
greater in the past, then the amount of dust at radii appropriate for emission at 60\micron\
might well be larger than a simple extrapolation from the dust responsible for the
3 -- 20\micron\ emission.  This would describe the opposite situation from that proposed by
\citep{suh97} who suggested higher mass-loss rates in the recent past to explain observations
of a number of other similar carbon stars.  The radial location of dust emitting strongly at
60\micron\ is of order 1 to a few arcseconds from the star.  For an assumed distance of 2.25 kpc
and its measured outflow velocity of 22 km s$^{-1}$ (Groenewegen et al. 2002), this would correspond
to a time of order 500 years to a couple thousand years in the past for the proposed higher mass-loss
rate.  This time scale is comparable to the period of fluctuation seen by \citet{mauron00} for IRC+10216.
 
The fact that we have constrained the absolute value and radial dependence of the dust density 
with our observations means that we have constrained the recent dust mass-loss rate for
AFGL 5440 as well.  The dust mass-loss rate implied by our derived optical depth and radial
density dependence is \massloss$_{dust}$ = 6.5 $\times$ 10$^{-8}$ \msun$/$yr for
the optical constants of amorphous carbon used by DUSTY \citep{ivez99}.   Groenewegen et al. (2002)
computed dust and gas mass-loss rates individually for AFGL 5440 on the basis of the IRAS 60\micron\
flux for the dust, and millimeter CO observations for the gas.  They derived a gas mass-loss
rate of 3.1$\times 10^{-5}$ \msun$/$yr and dust mass-loss rate of 5.1$\times$10$^{-8}$ \msun$/$yr
implying a gas-to-dust mass ratio of 600.  Interestingly, their dust mass-loss rate is slightly lower than the value
we have derived in spite of their using the 60\micron\ IRAS flux for normalization.  This result
reinforces the overall uncertainties in model assumptions and absolute dust opacties, particularly at
longer infrared wavelengths.  In any case, a dust mass-loss rate of order 5 -- 10$\times10^{-8} $\msun/yr and gas
mass-loss rate a few hundred times larger are consistent with the data.

In summary, our observations have separately resolved the stellar photosphere and the inner edge of the
circumstellar dust shell around AFGL 5440.  We have strongly constrained the inner radius of the dust shell surrounding the star
as well as the near-infrared optical depth.  Our constraints together with the spectral energy
distribution suggest a dust density gradient consistent with that expected for radiatively driven
mass loss with the exception that the far-infrared flux may imply a recent decrease in mass-loss
rate from the time when the far-ir-emitting dust was ejected.

\section{Acknowledgments}

We thank a number of people and institutions that have supported this work since its earliest stages.
M. Simon greatly encouraged our efforts and provided many useful comments over the
course of this work.  He also provided the basic prediction software (developed by L. Cassar) that
we use for determining the times and elements of occultation events.  We also acknowledge illuminating
conversations with A. Richichi, S. Guilloteau, D. Evans, and R. E. Nather and very helpful
suggestions from two anonymous referees.  D. Wilson developed the
initial versions of a number of the reduction algorithms used in this work and provided a great
deal of assistance during the observations.  C. Young provided help in the intricacies of DUSTY. This project has
been supported by: internal McDonald Observatory funding, NASA Grant NAG5-10458, and NSF grant AST-0096626.
We also acknowledge extensive use of the NASA Astrophysics Data System and SIMBAD, and P. Harvey thanks
the University of Colorado's Center for Astrophysics and Space Astronomy for graciously hosting him
during a sabbatical while much of this paper was written.



\begin{deluxetable}{cccccccc}
\tablecolumns{7}
\tablecaption{DUSTY Models of AFGL 5440\label{modtbl}}
\tablewidth{0pt} 
\tablehead{
\colhead{Model}                &
\colhead{T$_{max}$} &
\colhead{$\rho \propto r^{-N}$ } &
\colhead{$\tau$}  &
\colhead{Amorph C}  &
\colhead{Graphite}  &
\colhead{SiC}  &
\colhead{Size Dist.}               \\
\colhead{}                &
\colhead{K} &
\colhead{} &
\colhead{@5\micron}  &
\colhead{\%}  &
\colhead{\%)}  &
\colhead{\%)}  &
\colhead{}                  
}

\startdata

 194& 950 & -2.0 &   0.41  &  40 & 50 &  10 & MRN$^1$ \cr

 200& 950 & -2.0 &   0.55  &  95 & 0 &  5 & KMH$^2$ \cr

 206& 850 & rad-flow$^3$ &   0.45  &  95 & 0 &  5 & KMH$^2$ \cr
\enddata

\tablecomments{$^1$\citet{mathis77} with a$_{min} = .005$\micron; a$_{max} = 0.25$\micron.\\
$^2$\citet{kim94} with a$_{min} = .005$\micron; a$_{max} = 0.2$\micron.\\
$^3$ Radiatively driven outflow computed by DUSTY as described by \citet{ivez96}}
\end{deluxetable}


\clearpage

\begin{thebibliography}{199}

\bibitem[Andersen, Liodl,\& Hofner(1999)]{ander99}
Andersen, A. C., Liodl, R. \& Hofner, S. 1999, A\&A, 349, 243
\bibitem[Biller et al.(2005)]{biller05}
Biller, B. A. et al. 2005, \apj, 620, 450
\bibitem[Egan \& Leung(1995)]{egan95}
Egan, M. P. \& Leung, C. M. 1995, \apj, 444, 251
\bibitem[Ferrarotti \& Gail(2006)]{ferr06}
Ferrarotti, A. S. \& Gail, H.-P. 2006, A\&A, 447, 553
\bibitem[Fong, Meixner \& Shah(2003)]{fong03}
Fong, D., Meixner, M. \& Shah, R. Y. 2003, \apj, 582, L39
\bibitem[Groenewegen et al.(2002)]{groen02}
Groenewegen, M. A. T., Sevenster, M., Spoon, H. W. W. \& P\'erez, I. 2002, A \& A, 390, 511
\bibitem[Guandalini et al.(2006)]{guand06}
Guandalini, R., Busso, M., Ciprini, S., Silvestro, G. \& Persi, P. 2006, A\&A, 445, 1069
\bibitem[Guglielmo et al.(1993)]{gug93}
Guglielmo, F. et al. 1993, A\&A Suppl., 99, 31
\bibitem[Harvey \& Wilson(2003)]{harv03}
Harvey, P. M. \& Wilson, D. 2003, Proc. SPIE, 4841, 355
\bibitem[Hofmann et al.(2001)]{hoff01}
Hofmann, K.-H., Blocker, T., Weigelt, G. \& Balega, Y. 2001, A\&A, 379, 529
\bibitem[Ivezi\'c \& Elitzur(1996)]{ivez96}
Ivezi\'c, Z. \& Elitzur, M. 1996 MNRAS, 279, 1019
\bibitem[Ivezi\'c, Nenkova \& Elitzur (1999)]{ivez99}
Ivezi\'c, Z., Nenkova, M. \& Elitzur, M. 1999, User Manual for DUSTY, University of Kentucky Internal Report, accessible at http://www.pa.uky.edu/~moshe/dusty
\bibitem[Jura(2004)]{jura04}
Jura, M. 2004, ASP Conf. Series, 309, 321
\bibitem[Kim, Martin \& Hendry(1994)]{kim94}
Kim, S. H., Martin, P. G. \& Hendry P. D. 1994, \apj, 422, 164
\bibitem[Kwok, Volk \& Bidelman(1997)]{kwok97}
Kwok, S., Volk, K. \& Bidelman, W. P. 1997, \apj, 112, 557
\bibitem[Le Bertre(1997)]{lebert97}
Le Bertre, T. 1997, A\&A, 324, 1059
\bibitem[Le Bertre, Gougeon \& Le Sidaner(1995)]{lebert95}
Le Bertre, T., Gougeon, S. \& Le Sidaner, P. 1995, A\&A, 299, 791
\bibitem[Lorenz-Martins et al.(2001)]{lorenz01}
Lorenz-Martins, S., de Ara\'ujo, F. X., Codina Landaberry, S. J., de Almeida, W. G. \& de Nader, R. V. 2001, A\&A, 367, 189
\bibitem[Mathis, Rumpl \& Nordsieck(1977)]{mathis77}
Mathis, J. S., Rumpl, W. \& Nordsieck, K. H. 1977, \apj, 217, 425
\bibitem[Mauron \& Huggins(2000)]{mauron00}
Mauron, N. \& Huggins, P. J. 2000, A\&A, 359, 707
\bibitem[Monnier(2003)]{mon03}
Monnier, J.D. 2003, Rep. Prog. Phys., 66, 789
\bibitem[Nather \& McCants(1970)]{nather70}
Nather, R. E. \& McCants, M. M. 1970, AJ, 75, 963
\bibitem[Olofsson et al.(1996)]{oloff96}
Olofsson, H., Bergman, P., Eriksson, K. \& Gustafsson, B. 1996, A\&A, 311, 587
\bibitem[Quirrenbach(2004)]{quir04}
Quirrenbach, A. 2004, Adv. Sp. Res., 34, 524
\bibitem[Richichi et al.(1995)]{rich95}
Richichi, A. et al. 1995, A\&A, 301, 439
\bibitem[Schr\"oder et al.(1998)]{schroder98}
Schr\"oder, K.-P., Winters, J. M., Arndt, T. U. \& Sedlmayr, E. 1998, A\&A, 335, L9
\bibitem[Schmidtke, P. C. et al. (1986)]{schmid86}
Schmidtke, P. C. et al. 1986, \aj, 91, 961
\bibitem[Scholz(2003)]{scholz03}
Scholz, M. 2003, Proc. SPIE, 4838, 163
\bibitem[Suh(1997)]{suh97}
Suh, K. Y. 1997, MNRAS, 289, 559
\bibitem[Thompson, Creech-Eakman \& van Belle(2003)]{thomp03}
Thompson, R.R., Creech-Eakman, M.J. \& van Belle, G.T. 2003, Proc. SPIE, 4838, 221
\bibitem[van Belle(1999)]{vanbelle99}
van Belle, G.T. 1999, PASP, 111, 1515
\bibitem[Volk \& Cohen(1989)]{volk89}
Volk, K. \& Cohen, M. 1989, AJ, 98, 931
\bibitem[Wallerstein \& Knapp(1998)]{walknapp98}
Wallerstein, G. \& Knapp, G. R. 1998, ARAA, 36, 369
\bibitem[Winters et al.(1997)]{winters97}
Winters, J. M., Fleischer, A. J., Le Bertre, T. \& Sedlmayr, E. 1997, A\&A, 326, 305
\bibitem[Zuckerman \& Dyck(1986)]{zuck86}
Zuckerman, B. \& Dyck, M. 1986, \apj, 311, 345

\end{thebibliography}
\end{document}